\documentclass[12pt]{article}
\usepackage{epsfig}

\begin{document}

\title{The Generalized Harry  Dym Equation.}
\author{Ziemowit Popowicz \\
 University of Wroc\l aw,  Institute of Theoretical Physics\\
 pl.M.Borna 9  50-205 Wroc\l aw Poland\\
 e-mail: ziemek@ift.uni.wroc.pl }
\maketitle

to appear in Phys.Lett.A.

\begin{abstract}
The Harry Dym equation is generalized to the system of equations in the same manner as  the 
Korteweg - de Vries equation is generalized to the Hirota - Satsuma equation. The Lax and Hamiltonian 
formulation for this generalization is given. This  generalized Lax operator gives the hierarchy 
of equations also.

\end{abstract}

\vspace{2cm}

\section*{Introduction.} 
Large classes of nonlinear partial differential equations are integrable by the inverse spectral transform method 
and its modifications [1,2]. It is possible to construct such equations using the so called  Lax representation 
 
\begin{equation}
L:=\sum_iu_i(x)\partial^i,
\end{equation}
\begin{equation}
\frac{\partial L}{\partial t} = \Big [P_{\geq k}(L^q),L\Big ],
\end{equation}
where $k=0,1,2$ and $q$ may assume suitable integer or rational values.
By $P_{\geq k}$ we understand the projection 
to the sub-algebra of the pseudo differential symbols as discussed 
by Adler [3]
\begin{equation}
P_{\geq k}=\{ \sum_{i=k}^{\infty} a(x)_i\partial^{i} \}.
\end{equation}
The case $k=2$ corresponds to the Harry Dym hierarchy [1]. It is a huge class of equations. Therefore the 
problem to find  all or some particular reduction of this hierarchy seems to be very important. 

The Harry Dym equation has been recently generalized in different ways. 
An\-to\-nowicz and Fordy [4] introduced and investigated the coupled Harry Dym equations 
and showed that these 
equations are connected with the new isospectral flows. Different generalization of the Harry Dym equation 
have  been constructed in [5] as the bosonic sector of the fully supersymmetric $N=2$ extension of the 
Harry Dym hierarchy. The deformed Harry Dym equation has been considered recently in [6,7,8].

Beside these extensions we would like to present new  generalization of this hierarchy. 
We generalize Harry Dym equation in the same manner as the Korteweg de Vries equation is extended 
to the Hirota-Satsuma equation. This method could be considered as the admissible reduction of the
Harry Dym hierarchy. 

The paper is organized as follows. In the first section, we describe the Lax representation of 
the Hirota-Satsuma hierarchy and present its  modified version. The second section,  contains our basic construction  
where the Lax representation for the generalized Harry Dym equation is studied. In the next section,  
we show how it is possible to construct  the hierarchy for our generalization. In the  fourth section, 
we  show that, due to the constraints putting on the Lax operator for the modified Hirota-Satsuma equation, 
it is impossible to construct the reciprocal link to  the Lax operator for the generalized Harry Dym equation.
The last section, contains concluding remarks.

\section*{1. Hirota-Satsuma equation and its modification.}

The Hirota - Satsuma equation \cite{9}
\begin{eqnarray}
\frac{\partial u}{\partial t} &=&\Big (-u_{xxx}+3v_{xxx}-6u_xu+6vu_x+12v_xu\Big ), \\ \nonumber
\frac{\partial v}{\partial t} &=&\Big (-v_{xxx}+3u_{xxx}-6v_xv+6v_xu+12vu_x\Big ) ,
\end{eqnarray}
has the following Lax representation \cite{10}
\begin{equation}
\frac{\partial L}{\partial t} = 8\Big [(L^{3/4})_{\geq 0},L\Big ],
\end{equation}
\begin{equation}
L=\Big (\partial^2 + u\Big )\Big (\partial^2+v\Big ),
\end{equation}
where $\geq 0$ denotes purely differential part of $L^{3/4}$.

This Lax operator could be considered as the admissible reduction of the fourth-order Gelfand - Dikii 
Lax operator 
\begin{equation}
L:=f_4\partial^4+f_3\partial^3+f_2\partial^2 + f_1\partial + f_0,
\end{equation}
where 
\begin{equation}
f_4=1,\quad f_3=0,\quad f_2 = u+v, \quad f_1=2v_x, \quad f_0=v_{xx}+vu.
\end{equation}
It is possible to construct the hierarchy of system of equations using the Hirota-Satsuma Lax operator
\begin{equation}
\frac{\partial L}{\partial t_n} = 8\Big [(L^{(2n+1)/4})_{\geq 0},L\Big ],
\end{equation}
where $n=0,1,2...$.

Let us now apply the Miura transformation to $u,v$ 
\begin{equation}
u=-(g_x+g^2), \quad v=-(f_x+f^2)
\end{equation}
where $f$ and $g$ are the functions of $x$. One can easily check that, if $f$ and $g$ satisfy 
the following system of equations 
\begin{eqnarray}
\frac{\partial f}{\partial t} &=&\Big (-f_{xx}+2f^3+3g_{xx}-6g_xf+6g_xg-6g^2f\Big )_x ,\\ \nonumber
\frac{\partial g}{\partial t} &=&\Big (3f_{xx}+6f_xf-g_{xx}+2g^3-6gf_x-6gf^2\Big )_x ,
\end{eqnarray}
then $u$ and $v$ satsify the Hirota-Satsuma equation.
Therefore the system of equations (11)  could be considered as the modified Hirota-Satsuma equation 
which  possesses the following Lax representation
\begin{eqnarray}
\hat L &=& e^{(-\int dx \hspace{0.1cm} f)}Le^{(\int dx \hspace{0.1cm} f)} = \partial^4+4f\partial^3 + \\ \nonumber
&& (5f_x+5f^2-g_x-g^2)\partial^2 +(2f_{xx}+6f_xf+2f^3-2g_xf-2g^2f)\partial, 
\end{eqnarray}
\begin{equation}
 \frac{\partial \hat L}{\partial t}  =  8\Big [(\hat L^{3/4})_{\geq 1},\hat L\Big ].
\end{equation}

\section*{2. Generalized Harry Dym equation.} 

The Harry  Dym equation [1] 
\begin{equation}
\frac{\partial \omega}{\partial t} =\omega^3 \omega_{xxx}
\end{equation}
can be obtained from the following Lax representation  
\begin{equation}
L=\omega^2\partial^2, \label{harr}
\end{equation}
\begin{equation}
\frac{\partial L}{\partial t} = 4\Big [(L^{3/2})_{\geq 2},L\Big ].
\end{equation}

Let us now consider product of two different Lax operator (\ref{harr}) and define new Lax operator by 
\begin{equation}
L:= w^2\partial^2u^2\partial^2.
\end{equation}
Now the Lax representation 
\begin{equation}
\frac{\partial L}{\partial t} = 4\Big [(L^{3/4})_{\geq 2},L\Big ],
\end{equation}
leads  to the system of equations on $w$ and $u$
\begin{eqnarray}
\frac{\partial w}{\partial t} &=& w^3\Big (w^{-1/2}u^{3/2}\Big )_{xxx}, \\ \nonumber
\frac{\partial u}{\partial t} &=& u^3\Big (u^{-1/2}w^{3/2}\Big )_{xxx}.  \label{row}
\end{eqnarray}
It is our generalized Harry  Dym equation which can be written down in the Hamiltonian form as
\begin{equation}
\left(
\begin{array}{c}
w \\ u
\end{array}
\right)_t = J
\left (
\begin{array}{c} \frac{\delta H}{\delta w} \\ \frac{\delta H}{\delta u} \end{array} \right )=
\left(
\begin{array}{cc}
0 & w^3\partial^3 u^3  \\  u^3\partial^3 w^3 & 0 \end{array}\right) 
\left (
\begin{array}{c} \frac{\delta H}{\delta w} \\ \frac{\delta H}{\delta u} \end{array} \right ), 
\end{equation}
where 
\begin{equation}
H:=-2 \Big (wu\Big )^{-\frac{1}{2}}.
\end{equation}

\noindent Let us notice that our generalized Harry Dym equation (\ref{row}) can be rewritten as
\begin{equation}
\left(
\begin{array}{c}
f \\ g
\end{array}
\right)_t = \hat J
\left (
\begin{array}{c} \frac{\delta H}{\delta w} \\ \frac{\delta H}{\delta u} \end{array} \right )=
\left(
\begin{array}{cc}
0 & \partial^3   \\  \partial^3  & 0 \end{array}\right) 
\left (
\begin{array}{c} \frac{\delta H}{\delta f} \\ \frac{\delta H}{\delta g} \end{array} \right ) =
\left(
\begin{array}{c}
f^{{1}/{4}}g^{-{3}/{4}} \\ g^{{1}/{4}}f^{-{3}/{4}}
\end{array}
\right)_{xxx},
\end{equation}
where now $H=4(fg)^{1/4}$ and we 
transformed the  variables $w$ and $u$ to
\begin{equation}
w \rightarrow f^{-\frac{1}{2}}, \hspace{1cm} u \rightarrow g^{-\frac{1}{2}},
\end{equation}

The Hamiltonian operator $\hat J$  obviously satisfy the Jacobi identity. The transformation (23) 
could be considered as the Miura transformation between  $J$ and $\hat J$ operators. Due to it
the Hamiltonian operator $J$ satisfy the Jacobi identity too.

\section*{3. Generalized Harry Dym hierarchy.} 

The conserved charges for our generalized Harry Dym  equations could  be constructed 
as follows 

\begin{eqnarray}
H_1 &&= \int dx\hspace{0.2cm} tr L^{1/4} =\\ \nonumber 
&& - \int dx (5w_{xx}w^{-1/2}u^{1/2} + 2w_xu_xw^{-1/2}u^{-1/2} + 5u_{xx}u^{-1/2}w^{1/2})/16, 
\end{eqnarray}
\begin{eqnarray}
&& H_2  =  \int dx\hspace{0.2cm} tr L^{3/4} = \int dx  \\ \nonumber
&& \Big ( \Big (112w_{xx}^2w^{-1/2}u^{3/2} -
 56w_{xx}w_{x}^2w^{-3/2}u^{3/2} - 32w_{xx}w^{1/2}u_x^2u^{-1/2} + 7w_x^4w^{-5/2}u^{3/2} \\ \nonumber
&& \hspace{2cm}  - 144w_{xx}w^{1/2}u_{xx}u^{1/2} - 15w_x^2w^{-1/2}u_x^2u^{-1/2} \Big ) + 
 \Big ( w \rightleftharpoons u \Big )\Big ).
\end{eqnarray}

Similarly to the Hirota - Satsuma hierarchy our Lax operator (17) could be considered 
as the admissible reduction of the fourth-order Gelfand - Dikii Lax operator 
where now 
\begin{equation}
f_4=w^2u^2,\quad f_3=2w^2(u^2)_x,\quad f_2 = w^2(u^2)_{xx}, \quad f_1=0, \quad f_0=0.
\end{equation}
We can now construct the generalized hierarchy of Harry Dym equations considering the following 
Lax pair representation:
\begin{equation}
\frac{\partial L}{\partial t_n} = 4\Big [(L^{(2n+1)/4})_{\geq 2},L\Big ],
\end{equation}
where $n=0,1,2,...$.

The third member of  hierarchy is
\begin{eqnarray}
\frac{\partial w}{\partial t_3} &=& w^3\Big ( -12w_{xx}w^{-1/2}u^{5/2} +w_x^2w^{-3/2}u^{5/2} +\\ \nonumber 
&& \quad 10w_xw^{-1/2}u_xu^{3/2} +
  20w^{1/2}u_{xx}u^{3/2}-15w^{1/2}u_x^2u^{1/2} \Big )_{xxx}, \\ \nonumber
\frac{\partial u}{\partial t_3} &=& u^3\Big ( -12u_{xx}u^{-1/2}w^{5/2} +u_x^2u^{-3/2}w^{5/2} +\\ \nonumber 
&& \quad 10u_xu^{-1/2}w_xw^{3/2} +
  20u^{1/2}w_{xx}w^{3/2}-15u^{1/2}w_x^2w^{1/2}\Big )_{xxx}. 
\end{eqnarray}

This system of equations  is  Hamiltonian also with the same Hamiltonian operator $J$ as in (20). 
We can rewrtie these equations  as 
\begin{equation}
\left(
\begin{array}{c}
w \\ u
\end{array}
\right)_{t_3} = \left(
\begin{array}{cc}
0 & w^3\partial^3 u^3  \\  u^3\partial^3 w^3 & 0 \end{array}\right) 
\left (
\begin{array}{c} \frac{\delta H_1}{\delta w} \\ \frac{\delta H_1}{\delta u} \end{array} \right ).
\end{equation}

\section*{4. A reciprocal link. }

It is well known that the Harry Dym equation is connected with the Korteweg de Vries by the so called 
reciprocal auto-B\"{a}cklund transformation [11]. This transformation is realised in two steps.
In the first step, the  Korteweg de Vries equation is transformed to the modified Korteweg de Vries equation.
In the next, one performs the  transformation 
\begin{equation}
  \partial \Rightarrow \Phi_x\partial^{'} \Rightarrow \frac{\partial x^{'}}{\partial x} \frac{\partial}{\partial x^{'}},
 \hspace{1cm}    x^{'} \Rightarrow \Phi,
\end{equation}
for the Lax operator of the modified Korteweg de Vries equation.

Let us presents how the second step works for the  Harry Dym equation.
If we apply the transformation to the Lax operator responsible for the modified Korteweg de Vries equation
\begin{equation}
L=\partial^2 + 2v\partial,
\end{equation}
we obtain 
\begin{equation}
L^{'} = \Phi_{x}^{2}\partial_{x^{'}x^{'}}+(\Phi_{xx}+2\Phi_x v)\partial_{x^{'}},
\end{equation}
This operator is the Lax operator for the Harry Dym equation if 
\begin{equation}
w=\Phi_{x}, \quad v=-\frac{\Phi_{xx}}{2\Phi_x}
\end{equation}
It with the transformation (30) is desired reciprocal link to the Harry Dym Lax operator. 

Let us now try to apply the same strategy to our generalized Harry Dym equation. We have seen, 
in the first section, that the modified Hirota-Satsuma is described by tha Lax operator (12). If we 
use the $x$ transformation  (30) to this operator  we obtain 
\begin{eqnarray}
L^{'} &=& \Phi_x^4\partial^{'4}+\Big (6\Phi_{xx}\Phi_x^2+4\Phi_x^2 f\Big )\partial^{'3} \\ \nonumber 
&& \Big (4\Phi_{xxx}\Phi_x + 3\Phi_{xx}^2+12\Phi_{xx}\Phi_{x}+\Phi_{x}^2(5f_x+5f^2-g_x-g^2)\Big )\partial^{'2} \\ \nonumber
&& \Big (\Phi_{xxxx}+4\Phi_{xxx}f+\Phi_{xx}(5f_x+5f^2-g_x-g^2)+ \\ \nonumber
&& \Phi_x(2f_{xx}+6f_xf+2f^3-2g_xf-2g^2f) \Big )\partial^{'} .
\end{eqnarray}

We immediately realize that this operator can not coincides with the Lax operator (17) of the 
generalized Harry Dym equation. Indeed if these two operator are the same then we should assume 
that $\Phi_x=w^{1/2}u^{1/2}$. Then from the second term of (34)  we 
can fix  $f$ as a function of  $w$ and $u$. From the  the third terms of (34) we can fix $g_x+g^2$ and 
substitute to the last terms in (34). However this last term does not vanish and therefore we can not 
construct the reciprocal link to the Lax operator of the generalized Harry Dym eqaution. 

This is not in contradiction to the general construction of the reciprocal auto-B\"{a}cklund 
transformation for an arbitrary 
Gelfand-Dikii operator described in [11]. This general  construction  concerns to the unconstrained 
operators only. The Lax operator (12)  does not belong to this class. 
It contains two arbitrary functions but not three. These constraints 
does not allows to build  the reciprocal  transformation.

\section*{Conclusion.}
In this paper we generalized  the  Harry Dym equation to the system of equations in the same manner as  the 
Korteweg - de Vries equation is generalized to the Hirota - Satsuma equation. We presented 
the Lax and Hamiltonian formulation for this generalization. Moreover we showed that our generalization could be 
extended to the whole hierarchy of equations. We showed also that it is impossible to construct the 
reciprocal link for our Lax operator with the Lax operator of modified Hirota - Satsuma Lax operator.

Interestingly, the process of the generalizations of Harry Dym Lax operator to the product of several  
different Harry Dym operators, is restricted to two operators only. This can be easily verified by assuming the 
most general form on the solutions and checking the consistence of the Lax pair representation. 
The same observation concerns to the Korteveg de Vries operators.  

The system of equations considered in this paper seems to be new and needs further investigations. 
For example the problem of construction of the bihamiltonian structure and 
recursion operator seems to be very tempting.

\section*{Acknowledgments.} The author would like to thank to the anonymous referee for the valuable 
suggestions. Special thanks are to A.Fordy and S.Sakovich for the very stimulating discussion 
on the subject.


\begin{thebibliography}{99}

\bibitem{1} M.B\l aszak "{\it Multi - Hamiltonian Theory of Dynamical Systems}" Springer-Verlag 1998.

\bibitem{2} I.M.Gelfand, L.A.Dikii Funct.Anal.Appl. {\bf 10} (1976) 250.

\bibitem{3} M.Adler Invent Math. {\bf 50} (1979) 214.

\bibitem{4} M.Antonowicz, A.P.Fordy J.Phys.A:Math Gen {\bf 21} No.5 (1988) L269.

\bibitem{5} J.C.Brunelli, A.Das, Z.Popowicz \lq\lq {\it Supersymmetric extensions
 of the Harry Dym hierarchy} \rq\rq arXiv:nlin.SI/0304047 to be published in J.Math.Phys.

\bibitem{6} J.C.Brunelli, A.Das, Z.Popowicz \lq\lq {\it Deformed Harry Dym and Hunter-Zheng Equation} \rq\rq
arXiv:nlin.SI/0307043

 
\bibitem{7} M.S.Alber, R.Camassa, D.Holm, J.E.Marsden Proc. R.Soc.London {\bf 450A} (1995) 677. 

\bibitem{8} M.A.Manna, A.Neuve \lq\lq {\it A singular integrable equation from short capillary - gravity waves} \rq\rq
arXiv:physic/0303085.

\bibitem{9} R.Hirota and J.Satsuma Phys.Lett. {\bf 85A} (1981) 407.

\bibitem{10} R.Dodd and A.Fordy  Phys.Lett. {\bf 89A} (1982) 168.


\bibitem{11} B.G.Konopelchenko, W.Oevel Publ.Rims Kyoto Univ {\bf 29} (1993) 1.





\end{thebibliography}
\end{document}